\documentclass[
]{jss}

\usepackage[utf8]{inputenc}

\author{
Michael Kane\\Yale University \And Simon Urbanek\\The University of
Aukland
}
\title{On the Programmatic Generation of Reproducible Documents}

\Plainauthor{Michael Kane, Simon Urbanek}
\Plaintitle{On the Programmatic Generation of Reproducible Documents}
\Shorttitle{On the Programmatic Generation of Reproducible Documents}

\Abstract{
Reproducible document standards, like \proglang{R} Markdown, facilitate
the programmatic creation of documents whose content is itself
programmatically generated. While these documents are generally not
complete in the sense that they will not include prose content,
generated by an author to provide context, a narrative, etc.,
programmatic generation can provide substantial efficiencies for
structuring and constructing documents. This paper explores the
programmatic generation of reproducible by distinguishing components
than can be created by computational means from those requiring
human-generated prose, providing guidelines for the generation of these
documents, and identifying a use case in clinical trial reporting. These
concepts and use case are illustrated through the \pkg{listdown} package
for the \proglang{R} programming environment, which is is currently
available on the Comprehensive R Archive Network (CRAN).
}

\Plainkeywords{reproducibility, reproducible documents, data
presentation}

\Address{
    Michael Kane\\
  Yale University\\
  60 College StreetNew Haven, CT 06510, USA\\
  E-mail: \href{mailto:michael.kane@yale.edu}{\nolinkurl{michael.kane@yale.edu}}\\
  
      Simon Urbanek\\
  The University of Aukland\\
  38 Princes StreetAuckland Central, Aukland 1010, NZ\\
  E-mail: \href{mailto:s.urbanek@auckland.ac.nz}{\nolinkurl{s.urbanek@auckland.ac.nz}}\\
  
  }

\usepackage{amsmath}

\begin{document}

\hypertarget{background-and-concepts}{%
\section{Background and concepts}\label{background-and-concepts}}

\proglang{R} Markdown \citep{baumer2014} facilitates the construction of
computationally reproducible documents by allowing authors to insert
\proglang{R} code for data processing, exploration, analysis,
table-making, and visualization directly into structured, electronic
documents. The resulting documents are made up of these chunks of
\proglang{R} code, which we will refer to as \emph{computational
components} since they are generated by computational means, as well as
\emph{narrative components}, which (in scientific writing) is prose
intended to contextualize computational components, provide background,
define goals, establish themes, and convey the results of the document.
The integration of narrative and computational components was identified
as ``Literate Programming'' by \citet{knuth1984} and software tools,
like Sweave \citep{leisch2002}, have supported this functionality for
almost two decades.

\proglang{R} Markdown is particularly popular with its success likely
being driven by two factors. The first is the relative ease with which
these documents can be constructed. While \LaTeX ~is more expressive, it
is relatively technical and requires an investment in time to become
proficient. By contrast \proglang{R} Markdown documents are easier to
create and format and, when the document is used to create \LaTeX,
formatting can be passed through to the underlying .tex file. The second
factor driving adoption is likely its support for creating
\emph{modifiable} documents, namely Microsoft Word documents.
Researchers and analysts, especially those creating applied statistical
analyses, often collaborate with domain experts with less technical
knowledge. In these cases, the analyst focuses on creating the
computational components and narrative components related to results and
interpretation. After this initial document is created, the domain
expert is free to develop narrative components directly in the document
without needing to go through the analyst.

Since computational components are, by definition, computationally
derived objects and \proglang{R} Markdown is a well-defined standard, it
is possible to programmatically create \proglang{R} Markdown documents
with computational components, which is the focus of this paper.
Generating documents in this manner has two appealing characteristics.
First, it allows us to distinguish the presentation of analytical
results from other steps in a data science or data processing pipeline.
These other steps including cleaning and analysis often require their
own environment and configuration with requirements very different than
the computational needs of creating a presentation. By separating these
components each can be developed independently. At the same time, by
specifying a contract for the output of those objects, we can establish
a consistent means by which processed data can be passed to systems for
presentating those data in a structured way. The second reason for
programmatic creation of R Markdown documents is convenience. In
collaborative environments, especially in the early stages, large
numbers of graphs and tables are generated and discussed. By collecting
these artifacts and structuring them consistently, we can quickly
iterate upon and restructure the resulting documents to more clearly
present the data without needing to spend time on the creation of the
presentation document.

The \pkg{listdown} package provides functions to programmatically create
\proglang{R} Markdown files from named lists. It is intended for data
analysis pipelines where the presentation of the results is separated
from their creation. For this use case, a data processing (or analysis)
is performed and the results are provided in a single named list,
organized heirarchically. List element names denote sections,
subsections, subsubsection, etc. and the list elements contain the data
structure to be presented including graphs and tables. The package has
native support for \pkg{workflowr} \citep{blischak2019}, pdf, word, or
html document along with functions allowing a user to easily extend to
other types of supported documents. The goal of the package is to create
a documents with all tables and visualization that will appear
(computational components). This serves as a starting point from which a
user can organize outputs, describe a study, discuss results, and
provide conclusions (narrative components).

\pkg{listdown} therefore provides a reproducible means for producing a
document with specified computational components. It is most compatible
with data analysis pipelines where the data format is fixed but the
analyses are either being updated, which may affect narrative components
including the result discussion and conclusion, or where the experiment
is different, which affects all narrative components, but the data
format and processing is consistent. An example of the former is
provided later in this paper.

One area where we have found \pkg{listdown} is particuarly useful is in
the reporting and research of clinical trial data. These collaborations
tend to be between (bio)statisticians and clinicians either analyzing
past trial data to formulate a new trial or in trial monitoring where
trial telemetry (enrollment, responses, etc.) is reported and initial
analyses are conveyed to a clinician. The associated presentations
require very little context - clinicians often have a better
understanding of the data collected than the statistician - often
eliminating the need for narrative components. At the same time, a large
number of hierarchical, heterogenous artifacts (tables and multiple
types of plots) need to be conveyed thereby making the manual creation
of \proglang{R} Markdown documents inconvenient.

In this case, data presentation can be fixed across trials. This is
especially true in the initial stages, which focus on patient
demographics and enrollment. This approach has made it convenient for
our group to quickly generate standardize and complete reports for
multiple trials concurrently. To date, we have used listdown to report
on five clinical trials, with another two currently in process. Results
are disseminated using the \pkg{workflowr} package, usually with nine
tabs conveying aspects of the data from collection through several
different analyses, and each tab containing approxiately five to thirty
tables, plots, or other artifacts including \pkg{trelliscopejs}
\citep{trelliscopejs} environments which may hold hundreds of graphs. By
generating many presentation artifacts we are able to address data-drive
questions and issues during collaborative sessions and by carefully
structuring these elements we allowing all members to participate in the
process.

The \pkg{listdown} package itself is relatively simple with 6 distinct
methods that can be easily incorporated into existing analysis pipelines
for automatically creating documents that can be used for data
exploration and reviewing analysis results as well as a starting point
for a more formal write up. These methods include:

\begin{itemize}
\item{\bf as\_ld\_yml() }{- turn a computational component list into YAML with class information}
\item{\bf ld\_cc\_dendro() }{- create a dendrogram from a list of computational components}
\item{\bf ld\_chunk\_opts() }{- apply chunk options to a presentation object}
\item{\bf ld\_ioslides\_header() }{- create an ioslides presentation header}
\item{\bf ld\_make\_chunks() }{- write a listdown object to a string}
\item{\bf ld\_rmarkdown\_header() }{- create an R Markdown header}
\item{\bf ld\_workflowr\_header() }{- create a worflowr header}
\item{\bf listdown() }{- create a listdown object to create an R Markdown document}
\end{itemize}

The rest of this paper is structured as follow. The next section goes
over basic usage and commentary. This section is meant to convey the
basic approach used by the package and shows how to describe an output
document using \pkg{listdown}, create a document, and change how the
presentation of computational components can be specialized using
\pkg{listdown} decorators. With the user accustomed to the package's
basic usage, section 3 describes the design of the package. Section 4
goes over advanced usage of the package including adding initialization
code to and outputted document as well as how to control chunk-level
options. Section 5 provides a simplified case study of how the package
is currently being used in clinical trial reporting. Section 6 concludes
the paper with a few final remarks on the general types of applications
where \pkg{listdown} has been shown effective.

\hypertarget{basic-usage-and-overview}{%
\section{Basic usage and overview}\label{basic-usage-and-overview}}

Suppose we have just completed and analysis and have collected all of
the results into a list where the list elements are roughly in the order
we would like to present them in a document. It may be noted that this
is not always how computational components derived from data analyses
are collated. Often individual components are stored in multiple
locations on a single machine or across machines. However, it is
important to realize that even for analyses on large-scale data, the
digital artifacts to be presented are relatively small. Centralizing
them makes it easier to access them, since they don't need to be found
in mulitple locations. Also, storing them as a list provides a
hierarchical structure that translates directly to a document as we will
see below.

As a first example, we will consider the a list of visualizations from
the anscombe data set. The list is composed of four \pkg{ggplot2}
\citep{wickham2016}elements (named Linear, Non Linear, Outlier Vertical,
and Outlier Horizontal) each containing a scatter plot from the famous
Anscome Quartet - made available in the \pkg{datasets} package
\citep{R}. From the \texttt{computational\_components} list, we would
like to create a document with four sections with names corresponding to
the list names, each containing their respective visualizations.

\begin{CodeChunk}

\begin{CodeInput}
R> # Use ggplot2 to create the visualizations.
R> library(ggplot2)
R> 
R> # Load the Anscombe Quartet.
R> data(anscombe)
R> 
R> # Create the ggplot objects to display.
R>   computational_components <- list(
R+     Linear = ggplot(anscombe, aes(x = x1, y = y1)) + geom_point(),
R+     `Non Linear` = ggplot(anscombe, aes(x = x2, y = y2)) + geom_point(),
R+     `Outlier Vertical`= ggplot(anscombe, aes(x = x3, y = y3)) + 
R+       geom_point(),
R+     `Outlier Horizontal` =  ggplot(anscombe, aes(x = x4, y = y4)) + 
R+       geom_point())
R> 
R> # Save the file to disk to be read by the output R Markdown document.
R> saveRDS(computational_components, "comp-comp.rds")
\end{CodeInput}
\end{CodeChunk}

\hypertarget{creating-a-document-with-listdown}{%
\subsection{Creating a document with
listdown}\label{creating-a-document-with-listdown}}

Creating a document from the \texttt{computational\_components} requires
two steps. First, we will create a \texttt{listdown} object that
specifies how the \texttt{computational\_components} object will be
loaded into the document, which libraries and code needs to be included,
and how the list elements will be presented in the output R markdown
document.

\begin{CodeChunk}

\begin{CodeInput}
R> library(listdown)
R> 
R> ld <- listdown(load_cc_expr = readRDS("comp-comp.rds"),
R+                package = "ggplot2")
\end{CodeInput}
\end{CodeChunk}

The \texttt{ld} object, along with the computational components in the
\texttt{comp-comp.rds} file are sufficient to to create the sections,
subsections, and R chunks of a document. The only other thing requires
to create the document is the header. The listdown package currently
supports regular R Markdown and \pkg{workflowr} as \code{yml} objects
from the \pkg{yaml} package \citep{yaml}. These objects are stored as
named lists in R and are easily modified to accomodate document
parameters. A complete document can then be written to the console using
the code shown below. It could easily be written to a file for rendering
using the \texttt{writeLines()} function.

\begin{CodeChunk}

\begin{CodeInput}
R> doc <- c(
R+   as.character(ld_rmarkdown_header("Anscombe's Quartet",
R+                                    author = "Francis Anscombe",
R+                                    date = "1973")),
R+   ld_make_chunks(ld))
R> 
R> doc
\end{CodeInput}

\begin{CodeOutput}
 [1] "---"                                  
 [2] "title: Anscombe's Quartet"            
 [3] "author: Francis Anscombe"             
 [4] "date: '1973'"                         
 [5] "output: html_document"                
 [6] "---"                                  
 [7] ""                                     
 [8] "```{r}"                               
 [9] "library(ggplot2)"                     
[10] ""                                     
[11] "cc_list <- readRDS(\"comp-comp.rds\")"
[12] "```"                                  
[13] ""                                     
[14] "# Linear"                             
[15] ""                                     
[16] "```{r}"                               
[17] "cc_list$Linear"                       
[18] "```"                                  
[19] ""                                     
[20] "# Non Linear"                         
[21] ""                                     
[22] "```{r}"                               
[23] "cc_list$`Non Linear`"                 
[24] "```"                                  
[25] ""                                     
[26] "# Outlier Vertical"                   
[27] ""                                     
[28] "```{r}"                               
[29] "cc_list$`Outlier Vertical`"           
[30] "```"                                  
[31] ""                                     
[32] "# Outlier Horizontal"                 
[33] ""                                     
[34] "```{r}"                               
[35] "cc_list$`Outlier Horizontal`"         
[36] "```"                                  
\end{CodeOutput}
\end{CodeChunk}

The \texttt{listdown()} function provides \emph{document-wide} R chunk
options for displaying computational components. The chunk options are
exactly the same as those in the R markdown document and can be used to
tailor the default presentation for a variety of needs. The complete set
of options can be found in the R Markdown Reference Guide
\citep{rmarkdownref}. As a concrete example, the code used to create
present the plots could be hidden in the output document using the
following code.

\begin{CodeChunk}

\begin{CodeInput}
R> ld <- listdown(load_cc_expr = readRDS("comp-comp.rds"), 
R+                package = "ggplot2",
R+                echo = FALSE)
R> 
R> #cat(paste(ld_make_chunks(ld)[1:7], collapse = "\n"))
R> ld_make_chunks(ld)[1:7]
\end{CodeInput}

\begin{CodeOutput}
[1] ""                                     
[2] "```{r echo = FALSE}"                  
[3] "library(ggplot2)"                     
[4] ""                                     
[5] "cc_list <- readRDS(\"comp-comp.rds\")"
[6] "```"                                  
[7] ""                                     
\end{CodeOutput}
\end{CodeChunk}

\hypertarget{decorators}{%
\subsection{Decorators}\label{decorators}}

The first example is simple in part because the ggplot objects both
contains the data we want to display and, at the same time, provides the
mechanism for presenting them - rendering them in a graph. However, this
is not always the case. The objects being stored in the list of
computational components may not translate directly to the presentation
in a document. In these cases, a function is needed that takes the list
component and returns an object to be displayed. For example, suppose
that, along with showing graphs from the Anscombe Quartet, we would like
to include the data themselves. We could add the data to the
\texttt{computational\_components} list and then create the document
with:

\begin{CodeChunk}

\begin{CodeInput}
R> computational_components$Data <- anscombe
R> saveRDS(computational_components, "comp-comp.rds")
R> ld_make_chunks(ld)[32:36]
\end{CodeInput}

\begin{CodeOutput}
[1] "# Data"              
[2] ""                    
[3] "```{r echo = FALSE}"
[4] "cc_list$Data"        
[5] "```"                
\end{CodeOutput}
\end{CodeChunk}

In this case, the \pkg{listdown} package will show the entire data set
as is the default specified. However, suppose we do not want to show the
entire data set in the document. This is common, especially when the
data set is large and requires too much vertical space in the outputted
document resulting in too much or irrelevant data being shown. Instead,
we would like to output to an html document where the data is shown in a
\texttt{datatable} thereby controlling the amount of real-estate needed
to present the data and, at the same time, providing the user with
interactivity to sort and search the data set.

In \pkg{listdown}, a function or method that implements the presentation
of a computational component is referred to as a \emph{decorator} since
if follows the classic decorator pattern described in \citet{gamma1995}.
A decorator takes the element that will be presented as an argument and
returns an object for presentation in the output directory. A decorator
is specified using the \texttt{decorator} parameter of the
\texttt{listdown()} function using a named list where the name
corresponds to the type and the element correspond to the function or
method that will decorate an object of that type. For example, the
\texttt{anscombe} data set can be decorated with the
\texttt{DT::datatable()} function \citep{xie2020} as:

\begin{CodeChunk}

\begin{CodeInput}
R> ld <- listdown(load_cc_expr = readRDS("comp-comp.rds"), 
R+                package = c("ggplot2", "DT"),
R+                decorator = list(data.frame = datatable))
R> 
R> ld_make_chunks(ld)[33:37]
\end{CodeInput}

\begin{CodeOutput}
[1] "# Data"                  
[2] ""                       
[3] "```{r}"                  
[4] "datatable(cc_list$Data)"
[5] "```"                    
\end{CodeOutput}
\end{CodeChunk}

List names in the \texttt{decorator} argument provide a key to which a
function or method is mapped. The underlying decorator resolution is
implemented for a given computational component by going through
decorator names sequentially to see if the component inherits from the
name using the \texttt{inherits()} function. The function or method is
selected from the corresponding name which the element first inherits
from. This means that when customizing the presentation of objects that
inherit from a common class, the more abstract classes should appear at
the end of the list. This will ensure that specialized classes will be
encountered first in the resolution process. It should be noted that an
object's type is first checked against the decorator name list and then
checked to see if it is a list. This allows a user to both decorate a
list and retain \texttt{"list"} in its class attributes.

A separate argument, \texttt{default\_decorator}, allows the user to
specify the default decorator for an object whose type does not appear
in the \texttt{decorator} list. This allows the user to specify any
class name for the decorator and avoids a potential type name collision
with a default decorator whose name is determined by convention. By
default, this argument is set to \texttt{identity} but it can be use to
not display a computational component by default if the argument is set
to \texttt{NULL}.

\hypertarget{design}{%
\section{Design}\label{design}}

A \pkg{listdown} object specifies the location of a list of
computational components and options for presenting those components in
an R Markdown document. The list is a hierarchical data structure that
also provides the structure of the outputted document. A corresponding
document has two sections ``Iris'' and ``Sepal.Length''. The latter has
three subsections ``Sepal.Width'', ``Petal.Length'', and ``Colored''.
The ``Colored'' subsection has two sub-subsections, ``Sepal.Width'' and
``Petal.Length''. The structure can be seen using the
\texttt{ld\_cc\_dendro()} function.

\begin{CodeChunk}

\begin{CodeInput}
R> # Create a more hierarchical list of computational components.
R> comp_comp2 <- list(
R+   Iris = iris,
R+   Sepal.Length = list(
R+     Sepal.Width = ggplot(iris, aes(x = Sepal.Length, y = Sepal.Width)) + 
R+       geom_point(),
R+     Petal.Length = ggplot(iris, aes(x = Sepal.Length, y = Sepal.Width)) + 
R+       geom_point(),
R+     Colored = list(
R+       Sepal.Width = ggplot(iris, 
R+                           aes(x = Sepal.Length, y = Sepal.Width, 
R+                               color = Species)) + geom_point(),
R+       Petal.Length = ggplot(iris,
R+                             aes(x = Sepal.Length, y = Petal.Length, 
R+                                 color = Species)) + geom_point())))
R> 
R> # Create the dendrogram.
R> ld_cc_dendro(comp_comp2)
\end{CodeInput}

\begin{CodeOutput}

comp_comp2
  |-- Iris
  |  o-- object of type(s):data.frame
  o-- Sepal.Length
   |-- Sepal.Width
   |  o-- object of type(s):gg ggplot
   |-- Petal.Length
   |  o-- object of type(s):gg ggplot
   o-- Colored
    |-- Sepal.Width
    |  o-- object of type(s):gg ggplot
    o-- Petal.Length
       o-- object of type(s):gg ggplot
\end{CodeOutput}
\end{CodeChunk}

Both the \texttt{ld\_cc\_dendro()} and \texttt{ld\_make\_chunks()}
functions work by recursively descending the computational components
list depth-first. If the list containing and element has a name, it is
written to the output as a section, subsection, subsubsection, etc. to a
return string. If the visited list element is itself a list, then the
same procedure is called on the child list through a recursive call. If
the element is not a list, then it is outputted inside an R Markdown
chunk in the return string using the appropropriate decorator.

\hypertarget{advanced-usage}{%
\section{Advanced usage}\label{advanced-usage}}

\hypertarget{initialization-code}{%
\subsection{Initialization code}\label{initialization-code}}

The \texttt{listdown()} function facilitates the insertion of
initialization code through the \texttt{init\_expr} argument. Code is
inserted immediately after the libraries are loaded in the R Markdown
document. In general, it is suggested that the number of initial
expressions be kept small so that the \proglang{R} Markdown document is
easy to read. If a large number of functions are required by the target
\proglang{R} Markdown document then they can be put into a file and
sourced using the initial expression. As an example, suppose we are
creating an html document and presenting data using the
\texttt{datatable()} function. However, we do not want to include the
search capabilities provided. This can be easily accomplished by
creating a new function, \texttt{datatable\_no\_search()}, created using
the \texttt{partial()} function \cite{purrr} to partially apply
\texttt{list(dom\ =\ \textquotesingle{}t\textquotesingle{})} to the
\texttt{options} argument of \texttt{datatable}.

\begin{CodeChunk}

\begin{CodeInput}
R> saveRDS(comp_comp2, "comp-comp2.rds")
R> ld <- listdown(load_cc_expr = readRDS("comp-comp2.rds"),
R+                package = c("ggplot2", "DT", "purrr"),
R+                decorator = list(ggplot = identity,
R+                                 data.frame = datatable_no_search),
R+                init_expr = {
R+                  datatable_no_search <- partial(datatable,
R+                                                 options = list(dom = 't'))
R+                  },
R+                echo = FALSE)
R> 
R> ld_make_chunks(ld)[2:10]
\end{CodeInput}

\begin{CodeOutput}
[1] "```{r echo = FALSE}"                                                   
[2] "library(ggplot2)"                                                      
[3] "library(DT)"                                                           
[4] "library(purrr)"                                                        
[5] ""                                                                      
[6] "cc_list <- readRDS(\"comp-comp2.rds\")"                                
[7] ""                                                                      
[8] "datatable_no_search <- partial(datatable, options = list(dom = \"t\"))"
[9] "```"                                                                   
\end{CodeOutput}
\end{CodeChunk}

\hypertarget{r-code-chunk-customization}{%
\subsection{R code chunk
customization}\label{r-code-chunk-customization}}

The \pkg{listdown} package supports also supports capabilities to
further customize the presentation by specifying \proglang{R} code chunk
options in the \proglang{R} Markdown document in two distinct ways. The
first is used when the options we would like to specify is tied to type
of the object being presented. This can be though of as a chunk-option
decorator. The second is use for changing the options for an individual
chunk in an ad-hoc manner.

With three different modes of chunk customization it should be noted
that the increasing priority of the chunk options specification is
document-wide, decorator-wide, and ad hoc. That is, decorator-wide chunk
options take priority over document-wide chunk options and ad-hod
options take priority over decorator-wide options. In addition, it
should be noted that the use of the lowest priority scheme that
accomplishes the presentation goals is preferred because it lends itself
to greater code and maintenance efficiency.

\hypertarget{controlling-options-associated-with-a-decorator}{%
\subsection{Controlling options associated with a
decorator}\label{controlling-options-associated-with-a-decorator}}

The document-wide chunk option specification provides the default chunk
options for output documents generated using \pkg{listdown}. However,
the presentation of a data object often varies by type. For example, we
may want to specify the height and width of a graph, but not a table.
This is accomplished in the \pkg{listdown} package when a
\texttt{listdown} object is created using the
\texttt{decorator\_chunk\_opts} option in the \texttt{listdown()}
function. For example, associating all \texttt{ggplot} objects with
\proglang{R} chunks having a width of 100 and a height of 200 can be
accomplished with the following code and it can be seen that only chunk
options associated with a plot are modified.

\begin{CodeChunk}

\begin{CodeInput}
R> ld <- listdown(load_cc_expr = readRDS("comp-comp2.rds"),
R+                package = c("ggplot2", "DT", "purrr"),
R+                decorator_chunk_opts = 
R+                  list(ggplot = list(fig.width = 100,
R+                                     fig.height = 200)),
R+                init_expr = {
R+                  datatable_no_search <- partial(datatable,
R+                                                 options = list(dom = 't'))
R+                  },
R+                echo = FALSE)
R> 
R> ld_make_chunks(ld)[c(12:16, 19:24)]
\end{CodeInput}

\begin{CodeOutput}
 [1] "# Iris"                                                
 [2] ""                                                      
 [3] "```{r echo = FALSE}"                                   
 [4] "cc_list$Iris"                                          
 [5] "```"                                                   
 [6] ""                                                      
 [7] "## Sepal.Width"                                        
 [8] ""                                                      
 [9] "```{r echo = FALSE, fig.width = 100, fig.height = 200}"
[10] "cc_list$Sepal.Length$Sepal.Width"                      
[11] "```"                                                   
\end{CodeOutput}
\end{CodeChunk}

\hypertarget{controlling-chunk-level-options}{%
\subsection{Controlling chunk-level
options}\label{controlling-chunk-level-options}}

Along with providing decorator-wide chunk options, it is also possible
to control individual chunk options. The capability is distinct from the
document-wide and decorator-wide specification of options in that it
must be applied to the computational component list element whose
associated options will be modified, rather than the \texttt{listdown}
object. This is because the \texttt{listdown} only specifies how classes
of objects should be presented. To modify the chunk options associated
with a specific list element the list element is provided with a set of
attributes that can be queried by the \texttt{ld\_chunk\_opts()}
function as the output document is being generated. Because of the ad
hoc nature of this capability, its use is discouraged. A better
solution, that maintains the behavior is to add class information to the
list element and specify decorator-wide chunk options for the new class.
This maintains the separation of the computational component list, which
maintains the document structure and data for presentation from the
specification of how the document will be created and rendered.

\begin{CodeChunk}

\begin{CodeInput}
R> comp_comp2$Iris <- ld_chunk_opts(comp_comp2$Iris, echo = TRUE)
R> saveRDS(comp_comp2, "comp-comp2.rds")
R> ld_make_chunks(ld)[12:16]
\end{CodeInput}

\begin{CodeOutput}
[1] "# Iris"             
[2] ""                   
[3] "```{r echo = TRUE}"
[4] "cc_list$Iris"       
[5] "```"               
\end{CodeOutput}
\end{CodeChunk}

\hypertarget{a-simple-example-reporting-on-the-gtsummarytrial-dataset}{%
\section{A simple example: Reporting on the gtsummary::trial
dataset}\label{a-simple-example-reporting-on-the-gtsummarytrial-dataset}}

\label{sect:simple-example}

As mentioned before, we have found the \pkg{listdown} package
particularly helpful for reporting the results of clinical trials
thereby creating a basis for discussion and collaboration between
(bio)statisticians and the clinicians running the trials. For this use
case, the context and data collection procedures are well-understood and
as a result very few narrative components are needed. It is also the
case that the modes of presentation (tables, scatterplots,
survivalplots, consort diagrams, etc.) are standardized. The goal in
presenting the trial characteristics is to identify problems in the
data, monitor trial enrollment and response, quantify known
relationships among the data, and test hypotheses about a therapy's
efficacy.

In practice we generally separate the data cleaning, exploration,
analysis, monitoring, and presentation components. The computational
component list has tens of elements with thousands of visualizations.
These facts, coupled with the privacy constraints make a complete
example difficult for the purposes of this paper. So, below we provide a
simple example of the types of documents we provide using the
\texttt{gtsummary::trial} dataset. While it is not complete, it does
convey they types of reports we are currently generating for completed
and ongoing clinical trials.

The code creates a computational component list containing a table of
the patient characteristics along with survival plots by overall
survival, survival by stage, and survival by grade. The table is an
element of a named list called ``Table 1'' and then survival plots are
elements of named lists indicating the conditioning variable. The
structure can be seen in the dendrogram below.

\begin{CodeChunk}

\begin{CodeInput}
R> library(gtsummary)
R> library(dplyr)
R> library(survival)
R> library(survminer)
R> library(rmarkdown)
R> 
R> # Function to create the computational components.
R> make_surv_cc <- function(trial, treat, surv_cond_chars) {
R+   # Create Table 1 by treatment
R+   table_1 <- trial %>%
R+     tbl_summary(by = all_of(treat)) %>%
R+     gtsummary::as_flextable()
R+ 
R+   # Create survival plots base on survival characteristics
R+   scs <- lapply(c("1", surv_cond_chars),
R+                 function(sc) {
R+                   sprintf("Surv(ttdeath, death) ~ %s + %s", treat, sc) %>%
R+                     as.formula() %>%
R+                     surv_fit(trial) %>%
R+                     ggsurvplot()
R+                 })
R+   names(scs) <- c("Overall", tools::toTitleCase(surv_cond_chars))
R+   list(`Table 1` = table_1, `Survival Plots` = scs)
R+ }
R> 
R> # Create the computational components and write to surv-cc.rds.
R> surv_cc <- make_surv_cc(trial, treat = "trt",
R+                         surv_cond_chars = c("stage", "grade"))
R> 
R> ld_cc_dendro(surv_cc)
\end{CodeInput}

\begin{CodeOutput}

surv_cc
  |-- Table 1
  |  o-- object of type(s):flextable
  o-- Survival Plots
   |-- Overall
   |  o-- object of type(s):ggsurvplot ggsurv list
   |-- Stage
   |  o-- object of type(s):ggsurvplot ggsurv list
   o-- Grade
      o-- object of type(s):ggsurvplot ggsurv list
\end{CodeOutput}
\end{CodeChunk}

As shown before, the report is created by saving the computational
components, creating a listdown object, writing the \proglang{R}
Markdown document, and rendering it. The resulting document,
trial-report.html, can then be placed in a shared space where it can be
viewed and interpreted by stakeholders in the clincial trial. The
\proglang{R} Markdown document created by this code is shown in Appendix
1.

\begin{CodeChunk}

\begin{CodeInput}
R> # Turn the survplots into a list to show the individual components.
R> class(surv_cc$`Survival Plots`$Overall) <- 
R+   class(surv_cc$`Survival Plots`$Stage) <-
R+   class(surv_cc$`Survival Plots`$Grade) <- "list"
R> 
R> # Create a tab for each element of the survplot.
R> names(surv_cc$`Survival Plots`) <- 
R+   paste(names(surv_cc$`Survival Plots`), "{.tabset}")
R> 
R> # Rename the elements
R> names(surv_cc$`Survival Plots`$`Overall {.tabset}`) <- 
R+   names(surv_cc$`Survival Plots`$`Stage {.tabset}`) <- 
R+   names(surv_cc$`Survival Plots`$`Grade {.tabset}`) <- 
R+   c("Plot", "Data", "Table")
R>   
R> # Save the updated object for presentation.
R> saveRDS(surv_cc, "surv-cc.rds")
R> 
R> # Create the listdown object with appropriate dependencies
R> # and presentation parameters.
R> ld_surv <- listdown(load_cc_expr = readRDS("surv-cc.rds"),
R+                     package = c("gtsummary", "flextable", "DT", 
R+                                 "ggplot2"),
R+                     decorator_chunk_opts = 
R+                       list(gg = list(fig.width = 8,
R+                                      fig.height = 6)),
R+                     decorator = list(data.frame = datatable),
R+                     echo = FALSE,
R+                     message = FALSE,
R+                     warning = FALSE,
R+                     fig.width = 7,
R+                     fig.height = 4.5)
R> 
R> # Write an RMarkdown header and chunks derived from the 
R> # computational components and listdown object to trial-report.rmd.
R> writeLines(
R+   paste(c(
R+     as.character(ld_rmarkdown_header("Simple Trial Report")),
R+     ld_make_chunks(ld_surv))),
R+   "trial-report.rmd")
R> 
R> # Render the RMarkdown file to html and show it.
R> render("trial-report.rmd", quiet = TRUE)
R> browseURL("trial-report.html")
\end{CodeInput}
\end{CodeChunk}

\hypertarget{conclusion}{%
\section{Conclusion}\label{conclusion}}

While the programmtic generation of reproducible documents has appealing
qualities, it also has fundamental limitations that should be kept in
mind when deciding if tools like \pkg{listdown} should be employed.
First and foremost, without narrative components a document has very
little context. Quantitative analyses require research questions,
hypotheses, reviews, interpretations, and conclusions. Computational
components are necessary but generally not sufficient for constructing
an analysis. This means that if narrative components must be conveyed in
a document, then \pkg{listdown} may make the generation of their
presentation more convenient. However, it does not relieve the burden of
the author to create prose developing a narrative.

Second, it is difficult if not impossible to construct computational
components for an arbitary analyses. Analyses themselves have context
and are built with a set of assumptions and goals. Our experience shows
that \pkg{listdown} is easiest to used for a fixed data format. This
means a standard set of table and visualizations for similarly formatted
data. In particular, data that is periodicaly updated, without changes
to the format, is a case that is particularly amenable to document
generation reuse.

Keeping these limitations in mind, \pkg{listdown} can be used to
effectively reduce the the difficulty of generating documents in a
variety of contexts and fits readily in data processing and analysis
pipelines.

\bibliography{references}

\section*{Appendix 1: R Markdown produced in the simple example}
\label{appendix}

\begin{CodeChunk}

\begin{CodeInput}
R> ---
R> title: Simple Trial Report
R> output: html_document
R> ---
R> 
R> \```{r echo = FALSE, message = FALSE, warning = FALSE, fig.width = 7, 
R>      fig.height = 4.5}
R> library(gtsummary)
R> library(flextable)
R> library(DT)
R> library(ggplot2)
R> 
R> cc_list <- readRDS("surv-cc.rds")
R> \```
R> 
R> # Table 1
R> 
R> \```{r echo = FALSE, message = FALSE, warning = FALSE, fig.width = 7, 
R>      fig.height = 4.5}
R> cc_list$`Table 1`
R> \```
R> 
R> # Survival Plots
R> 
R> ## Overall {.tabset}
R> 
R> ### Plot
R> 
R> \```{r echo = FALSE, message = FALSE, warning = FALSE, fig.width = 8, 
R>      fig.height = 6}
R> cc_list$`Survival Plots`$`Overall {.tabset}`$Plot
R> \```
R> 
R> ### Data
R> 
R> \```{r echo = FALSE, message = FALSE, warning = FALSE, fig.width = 7, 
R>      fig.height = 4.5}
R> datatable(cc_list$`Survival Plots`$`Overall {.tabset}`$Data)
R> \```
R> 
R> ### Table
R> 
R> \```{r echo = FALSE, message = FALSE, warning = FALSE, fig.width = 7, 
R>      fig.height = 4.5}
R> datatable(cc_list$`Survival Plots`$`Overall {.tabset}`$Table)
R> \```
R> 
R> ## Stage {.tabset}
R> 
R> ### Plot
R> 
R> \```{r echo = FALSE, message = FALSE, warning = FALSE, fig.width = 8, 
R>      fig.height = 6}
R> cc_list$`Survival Plots`$`Stage {.tabset}`$Plot
R> \```
R> 
R> ### Data
R> 
R> \```{r echo = FALSE, message = FALSE, warning = FALSE, fig.width = 7, 
R>      fig.height = 4.5}
R> datatable(cc_list$`Survival Plots`$`Stage {.tabset}`$Data)
R> \```
R> 
R> ### Table
R> 
R> \```{r echo = FALSE, message = FALSE, warning = FALSE, fig.width = 7, 
R>      fig.height = 4.5}
R> datatable(cc_list$`Survival Plots`$`Stage {.tabset}`$Table)
R> \```
R> 
R> ## Grade {.tabset}
R> 
R> ### Plot
R> 
R> \```{r echo = FALSE, message = FALSE, warning = FALSE, fig.width = 8, 
R>      fig.height = 6}
R> cc_list$`Survival Plots`$`Grade {.tabset}`$Plot
R> \```
R> 
R> ### Data
R> 
R> \```{r echo = FALSE, message = FALSE, warning = FALSE, fig.width = 7, 
R>      fig.height = 4.5}
R> datatable(cc_list$`Survival Plots`$`Grade {.tabset}`$Data)
R> \```
R> 
R> ### Table
R> 
R> \```{r echo = FALSE, message = FALSE, warning = FALSE, fig.width = 7, 
R>      fig.height = 4.5}
R> datatable(cc_list$`Survival Plots`$`Grade {.tabset}`$Table)
R> \```
\end{CodeInput}
\end{CodeChunk}

\end{document}